\documentclass[fleqn,usenatbib]{mnras}
\usepackage[utf8]{inputenc}
\usepackage{mathrsfs}
\usepackage{natbib}
\usepackage{amsmath,amssymb}
\usepackage{amsfonts}
\usepackage{graphicx}
\usepackage{hyperref}
\usepackage[usenames,dvipsnames]{color}
\usepackage{mathrsfs}

\makeatletter

\providecommand{\tabularnewline}{\\}

\usepackage{mathrsfs}

\newcommand{\abs}[1]{\left\vert#1\right\vert}
\newcommand{\set}[1]{\left\{#1\right\}}

\makeatother

\title[G.W. Sources from Triple Interactions]{Analytic Modelling of Binary-Single Encounters: Non-Thermal Eccentricity Distribution and Gravitational-Wave Source Formation}

\author[Y. B. Ginat and H. B. Perets]{Yonadav Barry Ginat,$^{1}$\thanks{E-mail: ginat@campus.technion.ac.il}
and
Hagai B. Perets$^{1}$
\\
$^{1}$Faculty of Physics, Technion -- Israel Institute of Technology,
Haifa, 3200003, Israel
}

\date{Accepted XXX. Received YYY; in original form ZZZ}

\pubyear{2022}
\begin{document}
\label{firstpage}
\pagerange{\pageref{firstpage}--\pageref{lastpage}}
\maketitle

\begin{abstract}
Chaotic three-body interactions may lead to the formation of gravitational-wave sources. Here, by modelling the encounter as a series of close, non-hierarchical, triple approaches, interspersed with hierarchical phases, in which the system consists of an inner binary and a star that orbits it, we compute the pericentre probability distribution, and thereby the in-spiral probability in any given binary-single encounter. We then consider the indirect influence of binary-single encounters on the population of gravitational-wave sources, by changing the eccentricity distribution of hard binaries in clusters; we calculate this distribution analytically, by requiring that it be invariant under interactions with single stars.
\end{abstract}




\section{Introduction}

Chaotic three-body interactions are a dynamical pathway of gravitational-wave source production, especially in dense areas such as clusters
\citep{Hills1975,SigurdssinHernquist1993,QUINLAN1996,ZwartMcMillan2000,Rodriguezetal2015,Kocsis2020,Safarzadeh_2020,Tagawaetal2020,MandelBroakgaarden2022,Toonenetal2022}.
The rate of such phenomena is primarily determined by the probability that two of the three bodies approach each other sufficiently closely to coalesce within a dynamical time.  Previous progress on the probabilistic understanding  of the three-body problem is reviewed by\cite[][and references therein]{ValtonenKarttunen2006}; recently significant progress has been made \citep{StoneLeigh2019,Kol2021,GinatPerets2021a}, particularly advancing the understanding of the role played by the total angular momentum, which opened the door to analytical modelling of the probability of such interactions leading to gravitational-wave in-spirals.

In this paper, we are concerned only with encounters of single stars with hard binaries, which lead to so-called resonant interactions, in which there is a time when all three bodies are in an approximate energy equipartition.
As such an encounter proceeds as a series of hierarchical and non-hierarchical phases \citep{HutBahcall1983,Anosova1986,AnosovaOrlov1986,Samsingetal2014}, the probability of merging is determined by the probability distribution of the periapsis of the inner binary during a hierarchical phase, as well as the probability of close double approaches' occurring during a chaotic phase. As the initial binary is hard (i.e. its binding energy is much larger than the typical kinetic energy of a star in the cluster), it is necessarily the case that the interaction terminates when one of the three bodies is ejected to infinity \citep{Arnoldetal2006}, or by a merger or a collision.

Besides dynamically forming gravitational-wave sources, which are discussed in \S \ref{sec: gw in-spiral probability}, binary-single encounters also change the distribution of eccentricities in clusters: a binary's eccentricity typically changes considerably after each interaction with a single star; for an entire cluster, this process leads to a universal eccentricity distribution of hard binaries, which we endeavour to compute by studying the statistical solution of the non-hierarchical three-body problem in \S \ref{sec:multiple encounters}.
We begin, however, by discussing the pericentre distribution in \S \ref{sec:3d periapsis}, which we then use in \S \ref{sec: gw in-spiral probability} to derive the in-spiral probability.

\section{The Periapsis Probability Distribution}
\label{sec:3d periapsis}
The distribution of the orbital parameters of the remnant binary is \citep{GinatPerets2021a}
\begin{equation}
    f_b(E_b,\mathbf{S}|E,\mathbf{J})~\mathrm{d}^2\mathbf{S}\mathrm{d}E_b \propto \frac{m_b \theta_{\max}(E_b,R,\abs{\mathbf{J}-\mathbf{S}})~\mathrm{d}^2\mathbf{S}\mathrm{d}E_b}{\abs{\mathbf{J}-\mathbf{S}}\abs{E-E_b}^{3/2}\abs{E_b}^{3/2}}.
\end{equation}
Here, the binary energy is $E_b$, its angular momentum is $\mathbf{S}$, and its mass is $m_b$. The total, three-body energy is $E$ and the angular momentum is $\mathbf{J}$. We approximate $\theta_{\max}$ by $\theta_{ap}\sqrt{1-b^2}(1+2b^2)$, with $b$ defined in equation (A7) of \cite{GinatPerets2021a}. Quantities with a subscript $b$ refer to the remnant binary, and likewise a subscript $s$ refers to the orbit of the ejected star about this binary. Below we denote the total three-body mass by $M$. $R$ is the minimum distance the third body might have relative to the leftover binary's centre of mass, and still be in a hierarchical orbit; it is defined in equation (10) of \cite{GinatPerets2021a}.

Below we will mostly be concerned with the marginal distribution of $E_b$, and the binary's periapsis, $r_p$, chiefly at small values of $r_p$. This means that we should integrate $f_b$, multiplied by the Dirac delta function $\delta\left(r_p-a_b(1-e_b)\right)$, over $\mathbf{S}$; but first, let us derive the joint eccentricity and semi-major axis distribution. This is done by integrating over the direction of $\mathbf{S}$, i.e. over the inclination.  Let us define the auxiliary function
\begin{equation}
    \psi(x) = \frac{1}{4}\left(x \sqrt{1-x^2}\left(1+2x^2\right) + 3\arcsin x \right);
\end{equation}
integration over the direction of $\mathbf{S}$ yields,
\begin{equation}
    f_b(E_b,S|E,\mathbf{J})~\mathrm{d}S\mathrm{d}E_b \propto \frac{m_b \theta_{ap}(E_b,R)~\mathrm{d}S\mathrm{d}E_b}{\abs{E-E_b}^{3/2}\abs{E_b}^{3/2}}\times \frac{A_p}{J}\Delta \psi,
\end{equation}
where $A_p = \mu_s\sqrt{GMR}\sqrt{2\pm \frac{R}{a_s}}$ (the negative sign for ejection on a bound orbit, and \emph{vice versa}), and
\begin{equation}\label{eqn:delta psi definition}
    \Delta \psi \equiv \psi\left(\min\set{\frac{J+S}{A_p},1}\right) - \psi\left(\min\set{\frac{\abs{J-S}}{A_p},1}\right),
\end{equation}
provided that $J_m \leq J \leq J_M$, with
$J_m \equiv A_pj_m \equiv \max\set{0, S-\alpha}$ and $J_M \equiv A_pj_M \equiv \min\set{J_{\max},\alpha+S}$, where
\begin{equation}
    \alpha \equiv \begin{cases}
                    A_p, & \mbox{unbound} \\
                    \min\set{A_p,\mu_s\sqrt{GMR\eta\left(2-\eta \frac{R}{a_s}\right)}}, & \mbox{bound}.
                  \end{cases}
\end{equation}
At small $S \ll J, A_p$,
\begin{equation}\label{eqn:delta psi large J}
    \Delta\psi \approx 2S\sqrt{1-\frac{J^2}{A_p^2}}\left(1+2\frac{J^2}{A_p^2}\right).
\end{equation}
On the other hand, for small $S,J$, with $S > J$, one has $\Delta \psi \approx 2J/A_p$.

There are two possibilities: if $J$ is itself small, one expects a super-thermal eccentricity distribution \citep{StoneLeigh2019}, while for other values of $J$, it should become approximately thermal.
At any rate, the pericentre distribution is given by
\begin{equation}
\begin{aligned}
    f_b(r_p|E,\mathbf{J}) & = \iint f_b(E_b,S|E,\mathbf{J})\delta\left(r_p - a_b(1-e_b)\right)~\mathrm{d}S\mathrm{d}E_b \\ &
    \equiv \int f_b(r_p,E_b|E,\mathbf{J})~\mathrm{d}S.
\end{aligned}
\end{equation}

Let us approximate $f_b(r_p,E_b|E,\mathbf{J})$ for small values of $r_p$ (corresponding to high eccentricities, or small $S$), in the two afore-mentioned possibilities. First, if $J$ is not small relative to $A_p$, we find
\begin{equation}
\begin{aligned}
    f_{b,1}(r_p,E_b|E,\mathbf{J}) & \propto \frac{e_b}{\mu_b}\frac{\theta_{ap}J_c^2\Theta(a_b - r_p)}{J\sqrt{\abs{E_b}}\abs{E-E_b}^{3/2}} \\ & \times \sqrt{1-\frac{J^2}{A_p^2}}\left(1+2\frac{J^2}{A_p^2}\right),
\end{aligned}
\end{equation}
while for $J \ll A_p$, $S > J$,
\begin{equation}
\begin{aligned}
    f_{b,2}(r_p,E_b|E,\mathbf{J}) & \propto \frac{e_b}{\mu_b}\frac{\theta_{ap}J_c\Theta(a_b - r_p)}{\sqrt{\abs{E_b}}\abs{E-E_b}^{3/2}}\frac{2}{\sqrt{1-e_b^2}},
\end{aligned}
\end{equation}
where $e_b = 1 - r_p/a_b$.

\section{The In-Spiral Probability}
\label{sec: gw in-spiral probability}

The time to coalescence of a binary, due to gravitational-wave emission, is \citep[e.g.][and references therein]{Maggiore}
\begin{equation}
    \tau(a_b,e_b) = \tau_0(a_b)(1-e_b)^{7/2} g(e_b),
\end{equation}
where
\begin{equation}
    \tau_0(a_b) = \frac{5}{256}\frac{c^5a_b^4}{G^3m_b^2\mu_b}
\end{equation}
is the time to coalescence of a circular binary with initial semi-major axis $a_b$, and $g(e_b)$ is a slowly-varying function of $e_b$, that starts at $1$ at $e_b = 0$, and reaches $\frac{768}{425}$ at $e_b = 1$. Henceforth, we approximate $g(e_b) \approx 1$.

There are two possibilities for a gravitational-wave merger to be caused by a binary-single encounter:
\begin{enumerate}
    \item During a chaotic phase, one of the stars passes so close to another that some energy is radiated in gravitational waves, leaving the two as a soon-to-merge binary, while the third object is in a hierarchical orbit about it. The probability for this is \citep{HutInagaki1985,GinatPerets2021a} $p_{\rm gw,3b} = \frac{12a_0 r_0}{R^2}$, where $r_0$ is the maximum possible distance of closest approach between the two stars that would merge, for the merging to occur in a dynamical time, and $a_0$ is the original semi-major axis.
    \item One of the stars is ejected from the triple at the end of a chaotic phase, and the orbital parameters of the remnant binary yield a $\tau$ which is smaller than the orbital period of the outer binary (i.e. the inner binary merges before the third body returns).
\end{enumerate}

We therefore need to solve $\tau_{\rm dyn} \geq \tau(a_b, e_b)$, with
\begin{equation}
    \tau_{\rm dyn} \equiv 2\pi \frac{a_s^{3/2}}{\sqrt{GM}}.
\end{equation}
One can rearrange this inequality to the form $r_p \leq r_0(a_b)$, where
\begin{equation}
  r_0 \equiv a_b \min\set{1,\abs{1-\sqrt{1-\left(\frac{256G^3m_b^2\mu_b\tau_{\rm dyn}}{5c^5a_b^4}\right)^{2/7}}}}.
\end{equation}
Let $\mathcal{M}$ be the region in $(E_b,r_p)$-space where $\tau_{\rm dyn} \geq \tau$; because gravitational-wave emission is weak, $\mathcal{M}$ has a small probability. Furthermore, let us define $p_{\rm gw,bd} \equiv P(\mathcal{M}|\textrm{bound})$, i.e. the probability of having an inner binary whose orbital parameters satisfy $\tau_{\rm dyn} \geq \tau(a_b, e_b)$, given that the third body is ejected on a bound orbit. If we further neglect any losses of total energy and angular momentum (save for those occurring during a merger), then the random-walk model of \cite{GinatPerets2021a} predicts that the probability for a gravitational-wave merger to occur is
\begin{equation}\label{eqn: P GW}
\begin{aligned}
  P_{\rm gw} & = 1-\left[\sum_{n=1}^{\infty}p_{\rm ubd}p_{\rm bd}^{n-1}(1-p_{\rm gw,3b,ubd})\right.\\ &
  \times (1-p_{\rm gw,3b,bd})^{n-1}(1-p_{\rm gw,bd})^{n-1}\Bigg] \\ &
  = 1-\frac{p_{\rm ubd}(1-p_{\rm gw,3b,ubd})}{1-p_{\rm bd}(1-p_{\rm gw,bd})(1-p_{\rm gw,3b,bd})}.
\end{aligned}
\end{equation}
Here, $p_{\rm ubd} = 1-p_{\rm bd}$ is the probability of ejecting the third body to infinity, $p_{\rm gw,3b,ubd}$ is the afore-mentioned probability of having a gravitational-wave merger during the chaotic phase, leading to an unbound un-merged body (and similarly for $p_{\rm gw,3b,bd}$). We don't have a $p_{\rm gw,ubd}$, as once the encounter concludes with an ejection, we strictly do not consider any subsequent gravitational-wave merger of the remnant binary as one dynamically caused by a triple encounter.

\begin{figure}
    \centering
    \includegraphics[width=0.48\textwidth]{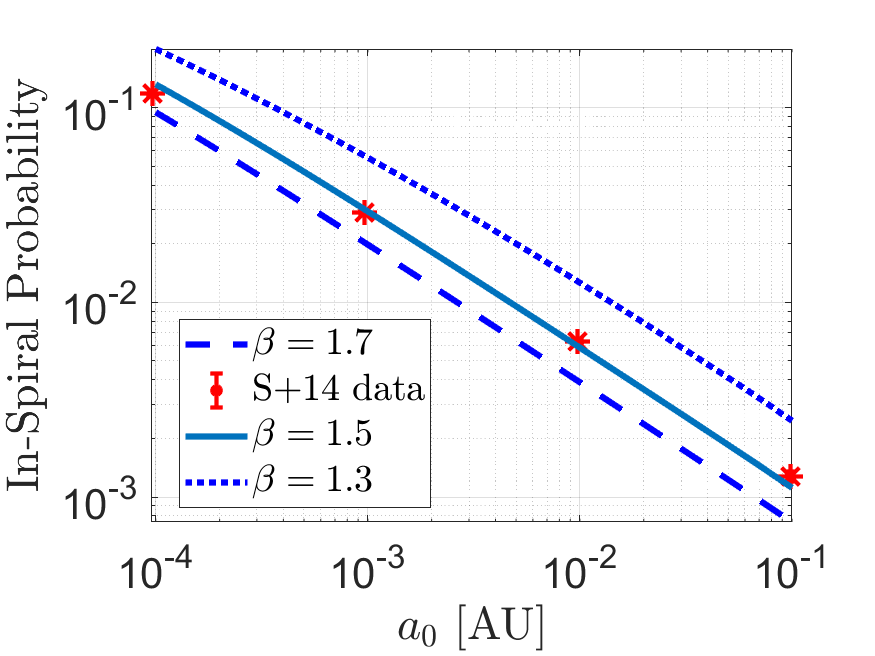}
    \caption{The gravitational-wave in-spiral/collision probability as a function of the initial semi-major axis. The particles are three $1~M_\odot$ black holes, and the in-coming star's velocity is $10~\textrm{km s}^{-1}$. The prediction of equation \eqref{eqn: P GW}, for $\eta = 5$ and various values of $\beta$, is plotted in blue and compared with the numerical results of \citet{Samsingetal2014} in red.}
    \label{fig:p gw 3d}
\end{figure}

\begin{figure}
  \centering
  \includegraphics[width=0.48\textwidth]{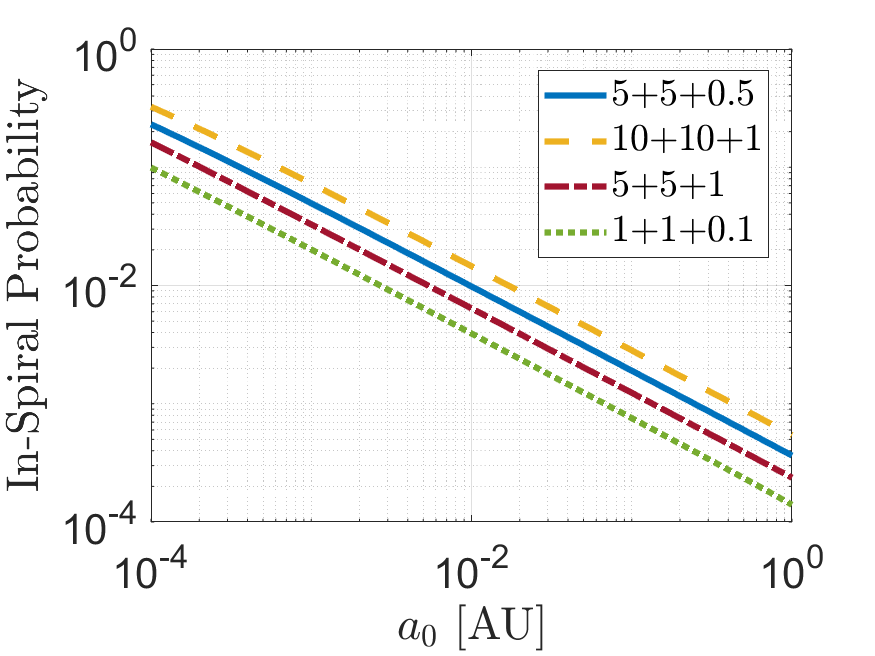}
  \caption{The same as figure \ref{fig:p gw 3d}, but with various masses.}\label{fig:p gw masses}
\end{figure}

If all the merging probabilities are very small, and if, barring gravitational-wave emission, there are $\langle N \rangle = \frac{1}{p_{\rm ubd}}$ chaotic phases until an ejection, the cross-section for a gravitational-wave merger in one of the two channels is
\begin{equation}\label{eqn:sigma gw 3d}
    \sigma_{\rm gw} \approx \langle N \rangle\left[12a_0\mathbb{E}\left(\frac{r_0(a_b)}{R^2}\right) + p_{\rm gw,bd}\right]\sigma_{\rm tot},
\end{equation}
$P_{\rm gw}$ is plotted in figure \ref{fig:p gw 3d}, and compared with the simulations of \cite{Samsingetal2014}. One can see that these results are in broad agreement with those of \cite{GinatPerets2021a}, who found $\beta \approx 1.3$, $\eta \approx 5$ to agree with the numerical data. The mild discrepancy between 1.3 and 1.5 is probably mostly due to the somewhat simplistic nature of the effective tidal model for energy losses adopted by that work, relative to the one incorporated in the numerical simulations, as well as the linear approximation (in the probability of significant tidal interactions, $12 a_0 r_{0}/R^2$) used there, which induces an error in $\beta$ of the order of $12r_0/(\beta a_0) \approx 0.25$. We use $\beta = 1.5$ and $\eta = 5$ in this paper, unless stated otherwise.
We also plot, in figure \ref{fig:p gw masses}, the gravitational-wave in-spiral probability for various mass configurations.

\begin{figure}
  \centering
  \includegraphics[width=0.48\textwidth]{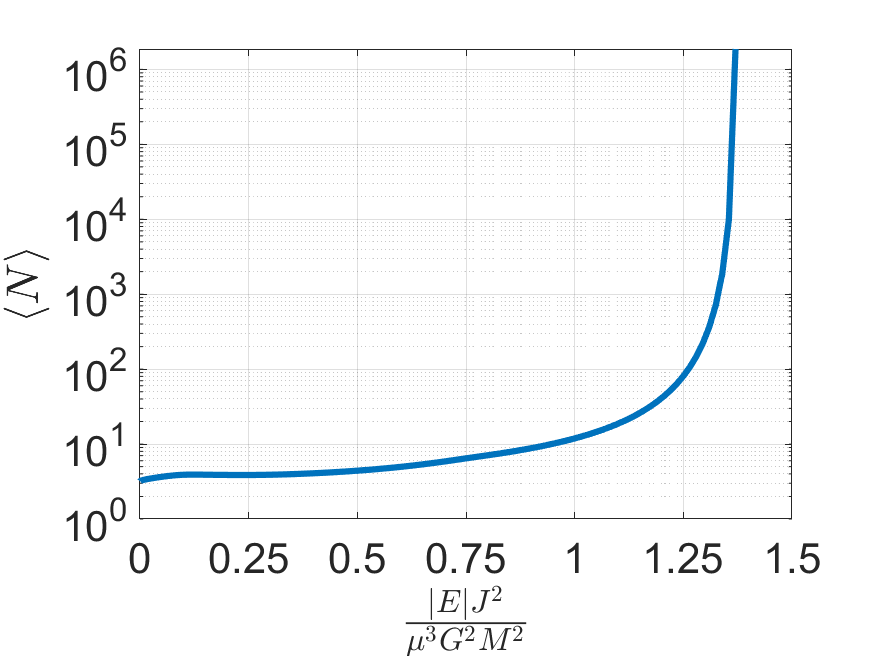}
  \caption{A plot of $\langle N \rangle$ as a function of the conserved quantities.}\label{fig:N of J}
\end{figure}

To conclude this section, we present a plot of $\langle N \rangle$ as a function of the initial conditions, for equal masses in figure \ref{fig:N of J}. Being dimension-less, it can only depend on the total energy and angular momentum in the combination $EJ^2/(\mu_s^3 G^2M^2)$ (and on mass-ratios), but it is extremely sensitive to the total angular momentum. Indeed, when the total angular momentum is too large, it is impossible for the system to be in the strong-interaction region of phase-space, and $f_b = 0$ \citep{StoneLeigh2019,GinatPerets2021a}. This is manifested by the probability for ejecting the third body to infinity on an unbound orbit occupying a smaller fraction of the probability mass, relative to the probability of ejecting it on a bound trajectory (which also decreases), as $J$ increases, until, at some value $J_*$, both vanish; in figure \ref{fig:N of J} this occurs at $\abs{E}J^2/(\mu_s^3G^2M^2) \approx 1.37$, but the validity of treating the encounter as a highly-chaotic one becomes moot already at $\abs{E}J^2/(\mu_s^3G^2M^2) \gtrsim 1.2$. Indeed, a numerical study by \cite{Parischewskyetal2021} found that, for two-dimensional scattering processes, the system is most chaotic when $J=0$. In future work we will use this computation of $\langle N \rangle$ as a function of the angular momentum to derive the life-time distribution of a resonant encounter.

Having computed the probability for one binary's interaction with one single star to produce a gravitational-wave in-spiral, we now turn to another possibility -- that interactions with many single stars will shape the eccentricity distribution of the entire cluster, in a way that may be favourable to the emergence of gravitational-wave sources.

\section{Multiple Encounters In Clusters}
\label{sec:multiple encounters}

After many encounters with single stars, one would expect the probability distribution of the eccentricities of binaries in a cluster to reach some steady-state distribution, which we denote by $f(E_b,\mathbf{S})$. Of course, the energy distribution of binaries does not, in general, reach a steady-state distribution, by Heggie's law \citep{Heggie1975,Binney}, but the eccentricity distribution does: after every encounter, the eccentricity effectively randomises \citep{StoneLeigh2019}, according to some distribution set by the total energy and angular momentum. If the remnant binary turns out to have a very high eccentricity, then it is probable that the eccentricity will be lower after the next one. The converse is also true, and therefore one expects the eccentricity distribution of many binaries in a cluster to tend towards some steady state. As the eccentricity varies considerably from one encounter to the next, this approach is much faster than the gradual hardening process, so one may treat the latter as adiabatic.
A different approach to describe the evolution of the distribution of orbital parameters of binaries in clusters is pursued by \cite{Leighetal2022}, with which the reader may compare our results.

The steady-state distribution must satisfy
\begin{equation}
    f(E_b,\mathbf{S}) = \int \mathrm{d}E \mathrm{d}^2\mathbf{J} f_b(E_b,\mathbf{S}|E,\mathbf{J}) p(E,\mathbf{J}),
\end{equation}
where $p(E,\mathbf{J})$ is the probability distribution of the total energy and angular momentum. For single stars that can come in from any direction, and at any impact parameter, the probability of finding $\mathbf{J}$ is approximately independent of the binary's previous angular momentum. Besides, as the binary is hard, $E$ is approximately equal to the binary's initial energy. Then,
\begin{equation}
    f(E_b,S) = \int \mathrm{d}E \iint_0^{J_{\max}(E)}\frac{J\mathrm{d}J\mathrm{d}\phi_J}{\pi J_{\max}^2(E)} f_b(E_b,S|E,\mathbf{J}) f(E),
\end{equation}
where $(J,\phi_J)$ are the polar co-ordinates of $\mathbf{J}$, and
\begin{equation}
    f(E) = \int \mathrm{d}S f(E,S).
\end{equation}
Assuming that $J_{\max}(E) + S \leq A_p$, we find that \begin{align}
    & \iint_0^{J_{\max}(E)}\frac{J\mathrm{d}J\mathrm{d}\phi_J}{\pi J_{\max}^2(E)} f_b(E_b,S|E,\mathbf{J}) \\ &
    \propto \frac{2A_p^2}{J_{\max}^2(E)} \frac{m_b \theta_{ap}(E,E_b,R)\Delta \phi}{\abs{E-E_b}^{3/2}\abs{E_b}^{3/2}},
\end{align}
where
\begin{equation}\label{eqn:delta_phi}
    \begin{aligned}
      & \Delta \phi(E,S) \equiv \Theta\left(1 - S - \frac{J_{\max}}{A_p}\right)\bigg[\phi(j_M + s) - \phi(s) + \phi(0) \\ & + \phi(\max\set{s-j_M,0}) - \phi(\max\set{j_M-s,0}) - \phi(s - j_m)\bigg] \\ &
      + \Theta(s-1)\bigg[\psi(1)(j_M - j_m) + \phi(0) + \phi(\max\set{s-j_M,0}) \\ &
       - \phi(\max\set{j_M-s,0}) - \phi(s - j_m)\bigg] \\ &
      + \Theta(1-s)\Theta\left(s + \frac{J_{\max}}{A_p} - 1\right)\bigg[\phi(1) - \phi(s) \\ & + \psi(1)(j_M-j_c) + \phi(0) + \phi(\max\set{s-j_M,0}) \\ &
      - \phi(\max\set{j_M-s,0}) - \phi(s - j_m)\bigg].
    \end{aligned}
\end{equation}
with $\phi(u)$ defined in equation (A12) of \cite{GinatPerets2021a}, $s \equiv S/A_p$, $j_M \equiv \min\set{J_{\max}/A_p,\alpha/A_p+s}$, $j_m \equiv \max\set{0, s-\alpha/A_p}$, and $\Theta$ denoting the Heaviside theta function.
Equation \eqref{eqn:delta_phi} is derived in appendix \ref{appendix:delta phi}. The maximum angular momentum is set by the requirement, that when the initial angular momenta of the in-coming star and the binary are aligned, the farthest distance of closest approach of the star and the centre-of-mass of the binary be equal to the latter's semi-major axis. This yields
\begin{equation}
    J_{\max} = \sqrt{2GM}\max \set{\mu_n \sqrt{R_n}}_{n=1}^3,
\end{equation}
where $n$ refers to the identity of the ejected particle, and $R_n$ is defined by equation (10) of \cite{GinatPerets2021a}, evaluated for the case where star $n$ is ejected, at $a_b$ which gives $E_b$ equal to the total three-body energy.

To find an analytical approximation of $f(E_b,S)$, we lastly assume that $f(E)$ is sharply peaked about some value $E_*$, so that eventually
\begin{equation}\label{eqn:distribution multiple}
    f(E_b,S) \propto \frac{2m_bA_p^2}{J_{\max}^2(E_*)} \frac{ \theta_{ap}(E_*,E_b,R)\Delta \phi(E_*,S)}{\abs{E_*-E_b}^{3/2}\abs{E_b}^{3/2}}.
\end{equation}
One requires additional information to determine $E_*$, \emph{viz.} the current mean binding energy of binaries in the cluster.

\begin{figure*}
  \centering
  \includegraphics[width=0.48\textwidth]{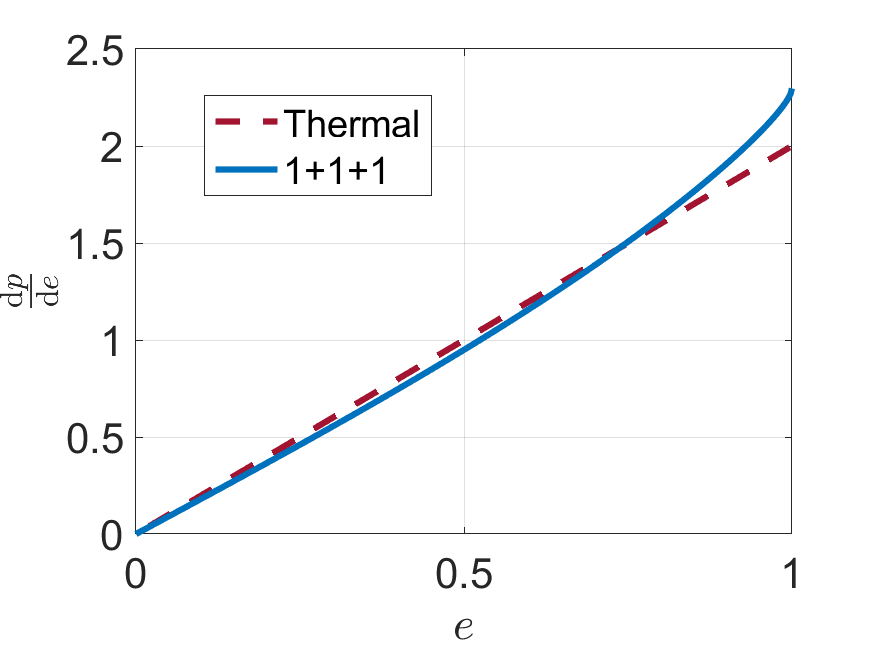}
  \includegraphics[width=0.48\textwidth]{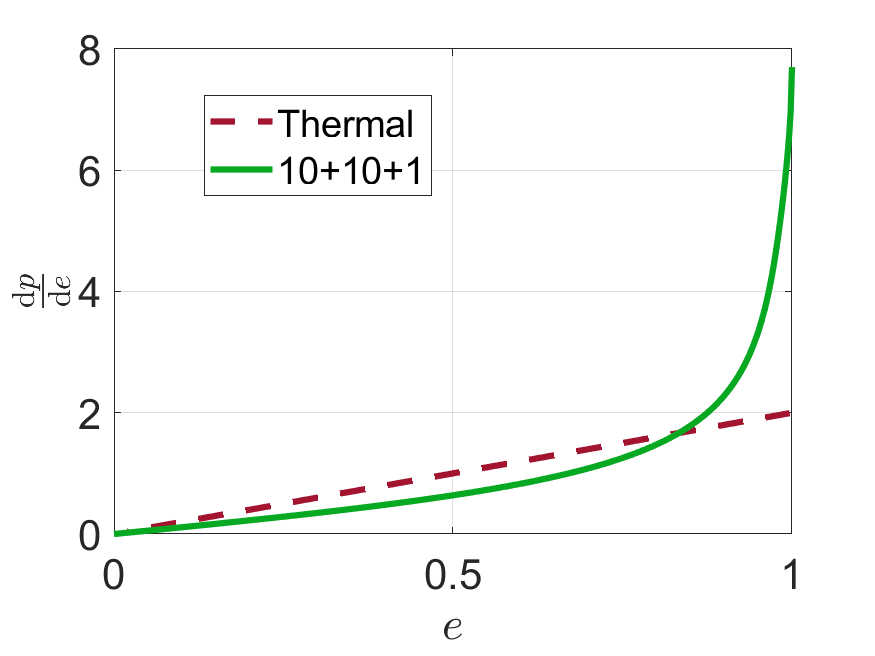}
  \caption{The marginal eccentricity distribution (normalised) arising from integrating equation \eqref{eqn:distribution multiple} over energies numerically, and changing from $S$ to $e_b$, compared with a thermal distribution (dashed line). The three bodies' masses are equal on the left panel, while they are $10~M_\odot$ in the binary, with a single perturber of $1~M_\odot$, on the right panel.}\label{fig:multiple marginal eccentricity}
\end{figure*}

The difference between $f$ and $f_b$ arises chiefly because while the latter is constrained by the total angular momentum, the former isn't, so its expression is actually simpler than that of $f_b$.
The marginal probability distribution of $f$ is not thermal (and likewise for $f_b$, as shown by \citealt{StoneLeigh2019}); it is somewhat super-thermal, as can be seen from figure \ref{fig:multiple marginal eccentricity}. The reason that this distribution is super-thermal is that $\Delta \psi$ in equation \eqref{eqn:delta psi definition} can yield either a thermal distribution (for large $J$, equation \ref{eqn:delta psi large J}) or a super-thermal one (for small $J$). So, when integrating over all allowed values of $J$, one has a convex combination of (close to) thermal distributions and super-thermal ones, to various degrees, which results in \eqref{eqn:distribution multiple}.
\begin{figure}
  \centering
  \includegraphics[width=0.48\textwidth]{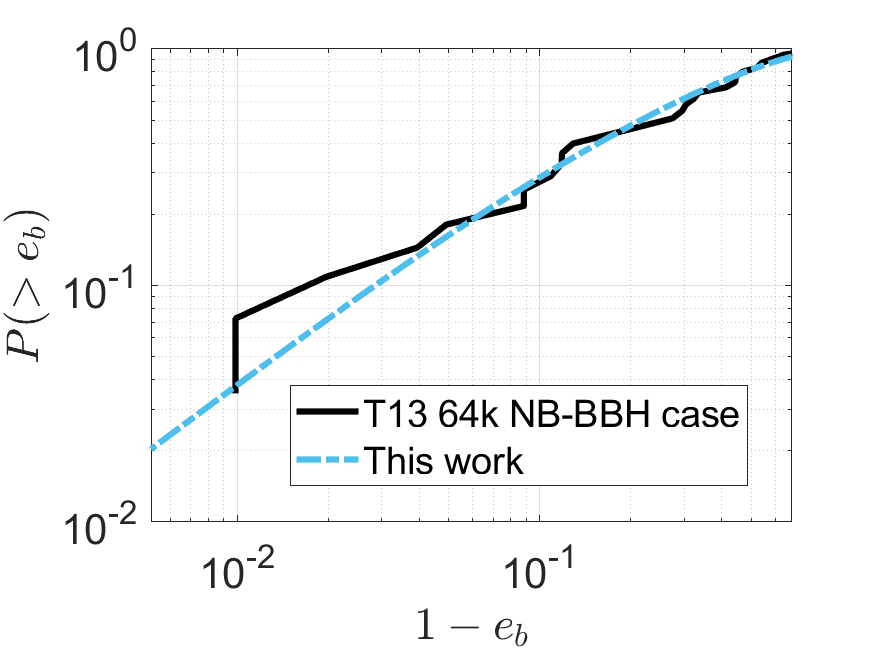}
  \caption{The marginal eccentricity distribution arising from integrating equation \eqref{eqn:distribution multiple} over energies numerically, and changing from $S$ to $e_b$, compared with numerical data from the $64$k case (largest number of particles) of figure 8 of \citet{Tanikawa2013}.}\label{fig:Tanikawa}
\end{figure}

Perhaps one might have expected $f$ to be a thermal distribution, based on the following (erroneous) reasoning: for a binary in thermal equilibrium, we know that the distribution of eccentricities is thermal. `Collisions' with single stars should allow for efficient energy and angular-momentum mixing, and therefore is expected to drive the distribution towards a thermal one. However, as we've shown here, these very encounters drive $f$ towards a distribution which is non-thermal; the reason is, of course, that (hard) binaries can never be in thermal equilibrium in a cluster that contains single stars, as already attested to by Heggie's law.

The results in this section, pertain to the single-mass case, where all stars in a cluster are of the same mass; of course, multi-mass multiple encounters are also possible in realistic clusters. Suppose, for example, that there are two types of astral objects in a cluster: a massive compact object, and a lighter star. Then, as time goes by, an ever increasing fraction of the binary population will be composed of two compact objects (as there is always an overwhelming probability to eject the lighter particle of the three; e.g. \citealt{Saslawetal1974,Hills1992,Heggieetal1996,GinatPerets2021a,Manwadkaretal2020,Kol2021,Manwadkaretal2021}), while the singles may contain both. If one assumes that this is the case, then one may use the {formul\ae} of \S \ref{sec:multiple encounters} straightforwardly. We show an example of this in figure \ref{fig:multiple marginal eccentricity}, which is different from the equal-mass case. In figure \ref{fig:Tanikawa} we compare these predictions to the numerical result of \cite{Tanikawa2013}, which considered the eccentricity distribution of tight black hole binaries that were dynamically ejected from a globular cluster, simulated by an N-body code. Our prediction that is plotted there is for $m_1 = m_2 = 10 ~M_\odot$ (close to the most likely binary component masses found there), with a perturber with mass $m \geq 1~ M_\odot$ distributed according to the mass function of \cite{Tanikawa2013}.\footnote{Perturbers with mass $m \lesssim 1 ~M_\odot$ were not included, since it is extremely unlikely that an interaction with a lighter star will eject the binary.} We expect the hard binaries that remain in the cluster to have a similar distribution after a sufficiently long time, and likewise for gravitational-wave source candidates. However, as shown by \cite{Gelleretal2019}, most binaries in clusters have not yet reached this stage, and their distribution is closer to the sub-thermal initial condition (see \citealt{DucheneKraus2013,Moe_2017} for observational evidence); besides, collisions, coupled to diffusion by weak encounters, can also act to deplete the high-eccentricity tail of the distribution (e.g. \citealt{Leighetal2022}).

\section{Conclusions}


The purpose of this paper is to provide an analytical formula for the probability distribution of the pericentre of a binary created by a close triple interaction, and to use it to compute the probability for a dynamically-generated gravitational-wave in-spiral, engendered in this way. The result is given by equation \eqref{eqn: P GW}. Here, we have been treating the three objects as point particles, but this is of course only true approximately. Indeed, at least when one of the particles is a star (i.e. not a compact object), collisions might be important (e.g. \citealt{HutInagaki1985,Daviesetal1993,Leighetal2017}); besides, here we considered only mergers, that occur during the three-body evolution, as dynamically-generated ones, but of course the final periapsis may be sufficiently small, that the remnant binary may merge before the next encounter \citep{Samsingetal2022}.

We have also shown how to compute the probability distribution of finding a hard binary with a given eccentricity, chosen at random from a cluster, when one takes into account multiple encounters with single stars. This distribution, equation \eqref{eqn:distribution multiple} was found to be somewhat super-thermal, which is important for estimating the over-all rates of gravitational-wave in-spirals \citep{Kocsis2020,MandelBroakgaarden2022}, especially in gas-assisted mergers \citep{RoznerPerets2022}.

Observational evidence seems to suggests an eccentricity distribution of binaries which is shallower than the thermal one \cite{DucheneKraus2013,Moe_2017}; an intriguing possibility is that it is this discrepancy originates from a mass-and-number-weighted superposition of the distributions shown in figure \ref{fig:multiple marginal eccentricity} and described in \S \ref{sec:multiple encounters}.

\section*{Acknowledgements}
We wish to thank Nathan Leigh for helpful comments. Y.B.G. is grateful for the support of the Adams Fellowship Programme of the Israeli Academy of Sciences and Humanities.

\section*{Data Availability}
The data underlying this article will be shared on reasonable request to the corresponding author.




\bibliographystyle{mnras}
\bibliography{encounters}



\appendix
\section{Derivation of $\Delta \phi$}
\label{appendix:delta phi}
To integrate over the angular momentum, we need to integrate equation \eqref{eqn:delta psi definition} over $J$, from  $J_m \equiv A_pj_m = \max\set{0, S-\alpha}$ to $J_M \equiv A_pj_M \equiv \min\set{J_{\max},\alpha+S}$. Note, that $S-J_m \leq A_p$, and since $J_m \leq S$, $\abs{S-J_m}\leq A_p$. In \S \ref{sec:multiple encounters} we only need the unbound case. As $J$ increases from $J_m$ to $S$, this absolute value decreases to $0$. Then, from $J = S$ to $J_M$, $\abs{J-S}$ increases from $0$ to $J_M - S \leq A_p$. Consequently,
\begin{equation}
    \min\set{\frac{\abs{J-S}}{A_p},1} = \frac{\abs{J-S}}{A_p}
\end{equation}
over the entire allowed range of $J$.
There are now three cases: $S>A_p$, $S\leq A_p$ but $S+J_{\max} > A_p$, and $S + J_{\max} \leq A_p$.

\subsection{Case $S>A_p$}
Equation \eqref{eqn:delta psi definition} reduces to
\begin{equation}
    \psi(1) - \psi(\abs{j-s}),
\end{equation}
where $j \equiv J/A_p$. We need to integrate
\begin{align}
    \Delta \phi & = \int_{j_m}^{j_M} \left[\psi(1) - \psi(\abs{j-s})\right]\mathrm{d}j \\ &
    = \psi(1)(j_M - j_m) - \int_{j_m}^{\min\set{s,j_M}}\psi(s-j)\mathrm{d}j \\ &
    - \int_s^{\max\set{j_M,s}}\psi(j-s)\mathrm{d}j \\ &
    = \psi(1)(j_M - j_m) + \phi(0) - \phi(s-j_m) \\ &
    - \phi(\max\set{j_M-s,0}) + \phi(\max\set{0,s-j_M}).
\end{align}

\subsection{Case $S \leq A_p$ But $S+J_{\max} > A_p$}
Now equation \eqref{eqn:delta psi definition} yields
\begin{equation}
    \psi(\min\set{s+j,1}) - \psi(\abs{s-j}).
\end{equation}
Observe that if $S\leq A_p$, $J_m = 0$; on the other hand, it is also true that if $J_m + S < A_p$, then $S\leq A_p$, so $J_m = 0$.

Let $j_c \equiv 1-s$, whence by definition
\begin{equation}
    0 = j_m \leq j_c \leq J_{\max}/A_p \leq j_M.
\end{equation}
The second integral, of $\psi(\abs{s-j})$, is the same as above. The first is
\begin{align}
    & \int_{j_m}^{j_M}\psi(\min\set{s+j,1})\mathrm{d}j = \int_{0}^{j_M}\psi(\min\set{s+j,1})\mathrm{d}j \\ &
    = \int_0^{j_c}\psi(s+j)\mathrm{d}j + \psi(1)(j_M-j_c) \\ &
    = \phi(1) - \psi(s) + \psi(1)(j_M-j_c).
\end{align}
Together with the second term, we find
\begin{align}
    \Delta \phi & = \phi(1) - \psi(s) + \psi(1)(j_M-j_c) + \phi(0) - \phi(s-j_m) \\ &
    - \phi(\max\set{j_M-s,0}) + \phi(\max\set{0,s-j_M}).
\end{align}

\subsection{Case $S+J_{\max} \leq A_p$}
In this case, since $J_{\max} \leq A_p -S \leq A_p + S$, the upper limit is $J_M = J_{\max}$. As before, $J_m = 0$. Then it is always the case that $s+j \leq 1$, so
\begin{equation}
    \Delta \psi = \psi(s+j) - \psi(\abs{s-j}).
\end{equation}
Performing the integrations yields
\begin{align}
    \Delta \phi & = \phi(s+j_M) - \phi(s) + \psi(1)(j_M-j_c) + \phi(0) - \phi(s-j_m) \\ & - \phi(\max\set{j_M-s,0}) + \phi(\max\set{0,s-j_M}).
\end{align}


\bsp	
\label{lastpage}
\end{document}